\documentclass[twocolumn,superscriptaddress,nobibnotes]{revtex4}

\usepackage{graphicx}
\usepackage{dcolumn}
\usepackage{bm}
\usepackage{times,mathptmx}
\usepackage[rgb,dvipsnames]{xcolor}
\usepackage{multirow}
\usepackage{enumitem}
\usepackage{chngpage}
\usepackage{float} 

\usepackage{amsmath}

\begin{document}


\title{\bf Self-injection-locked second-harmonic integrated source}
\author{Jingwei Ling}
\thanks{These authors contributed equally.}
\affiliation{Department of Electrical and Computer Engineering, University of Rochester, Rochester, NY 14627, USA}
\author{Jeremy Staffa}
\thanks{These authors contributed equally.}
\affiliation{Institute of Optics, University of Rochester, Rochester, NY 14627, USA}
\author{Heming Wang}
\thanks{These authors contributed equally.}
\affiliation{T. J. Watson Laboratory of Applied Physics, California Institute of Technology, Pasadena, CA 91125, USA}
\author{Boqiang Shen}
\thanks{These authors contributed equally.}
\affiliation{T. J. Watson Laboratory of Applied Physics, California Institute of Technology, Pasadena, CA 91125, USA}
\author{Lin Chang}
\thanks{These authors contributed equally.}
\affiliation{Department of Electrical and Computer Engineering, University of California Santa Barbara, Santa Barbara, CA 93106, USA}
\author{Usman A. Javid}
\affiliation{Institute of Optics, University of Rochester, Rochester, NY 14627, USA}
\author{Lue Wu}
\affiliation{T. J. Watson Laboratory of Applied Physics, California Institute of Technology, Pasadena, CA 91125, USA}
\author{Zhiquan Yuan}
\affiliation{T. J. Watson Laboratory of Applied Physics, California Institute of Technology, Pasadena, CA 91125, USA}
\author{Raymond Lopez-Rios}
\affiliation{Institute of Optics, University of Rochester, Rochester, NY 14627, USA}
\author{Mingxiao Li}
\affiliation{Department of Electrical and Computer Engineering, University of Rochester, Rochester, NY 14627, USA}
\author{Yang He}
\affiliation{Department of Electrical and Computer Engineering, University of Rochester, Rochester, NY 14627, USA}
\author{Bohan Li}
\affiliation{T. J. Watson Laboratory of Applied Physics, California Institute of Technology, Pasadena, CA 91125, USA}
\author{John E. Bowers}
\email{jbowers@ucsb.edu}
\affiliation{Department of Electrical and Computer Engineering, University of California Santa Barbara, Santa Barbara, CA 93106, USA}
\author{Kerry J. Vahala}
\email{vahala@caltech.edu}
\affiliation{T. J. Watson Laboratory of Applied Physics, California Institute of Technology, Pasadena, CA 91125, USA}
\author{Qiang Lin}
\email{qiang.lin@rochester.edu}
\affiliation{Department of Electrical and Computer Engineering, University of Rochester, Rochester, NY 14627, USA}
\affiliation{Institute of Optics, University of Rochester, Rochester, NY 14627, USA}



\begin{abstract}
High coherence visible and near-visible laser sources are centrally important to the operation of advanced position/navigation/timing systems \cite{lutwak2014micro} as well as classical/quantum sensing systems \cite{degen2017quantum}. However, the complexity and size of these bench-top lasers is an impediment to their transitioning beyond the laboratory. Here, a system-on-a-chip that emits high-coherence visible and near-visible lightwaves is demonstrated. The devices rely upon a new approach wherein wavelength conversion and coherence increase by self-injection-locking are combined within in a single nonlinear resonator. This simplified approach is demonstrated in a hybridly-integrated device and provides a short-term linewidth around 10-30~kHz. On-chip, converted optical power over 2 mW is also obtained. Moreover, measurements show that heterogeneous integration can result in conversion efficiency higher than 25\% with output power over 11~mW. Because the approach uses mature III-V pump lasers in combination with thin-film lithium niobate, it can be scaled for low-cost manufacturing of high-coherence visible emitters. Also, the coherence generation process can be transferred to other frequency conversion processes including optical parametric oscillation, sum/difference frequency generation, and third-harmonic generation.

\end{abstract}

\maketitle

\noindent\rmfamily\textbf{Introduction}\nolinebreak

\noindent\rmfamily



Optical frequency conversion based upon a quadratic optical nonlinearity is a powerful technology to transfer high-coherence laser radiation into new frequencies \cite{fejer1994nonlinear, dunn1999parametric, ebrahim2010continuous}.  And, in recent years the development of nonlinear photonic integrated circuits (PICs), particularly using the on-chip lithium-niobate-on-insulator (LNOI) platform \cite{boes2018status, zhu2021integrated, chang2016thin, wolf2018quasi, wang2018ultrahigh, luo2019optical, lin2019broadband, rao2019actively, chen2019ultra,  lu2019periodically, niu2020optimizing, zhao2020shallow, lu2020toward, mckenna2021ultra, chen2021efficient, wang2021efficient}, have boosted nonlinear conversion efficiency while enabling photonic integration with active and passive waveguide elements. However, to achieve high coherence in these systems bench top source lasers have been used. Here we demonstrate for the first time, to the best of our knowledge, a hybridly-integrated laser that produces efficient and ultra-coherent visible and near-visible light. The device combines second harmonic generation (SHG) in an LNOI microresonator that also functions to line narrow a DFB pumping laser through self-injection-locking (SIL) \cite{dahmani1987frequency,li1989analysis,hjelme1991semiconductor,jin2021hertz,li2021reaching}. This means that line narrowing and resonant enhancement for harmonic generation occur simultaneously, and also without need for servo-control of the pumping laser for frequency tracking to the nonlinear resonator.  The simultaneous nature of the harmonic and line-narrowing co-functions of the resonator is verified by measurements showing that frequency noise spectra of the line-narrowed pump light and the harmonic emission precisely match, but with an offset of 6 dB that is determined by the physics of the second-harmonic process.

\begin{figure*}
	\centering
	\includegraphics[width=\linewidth]{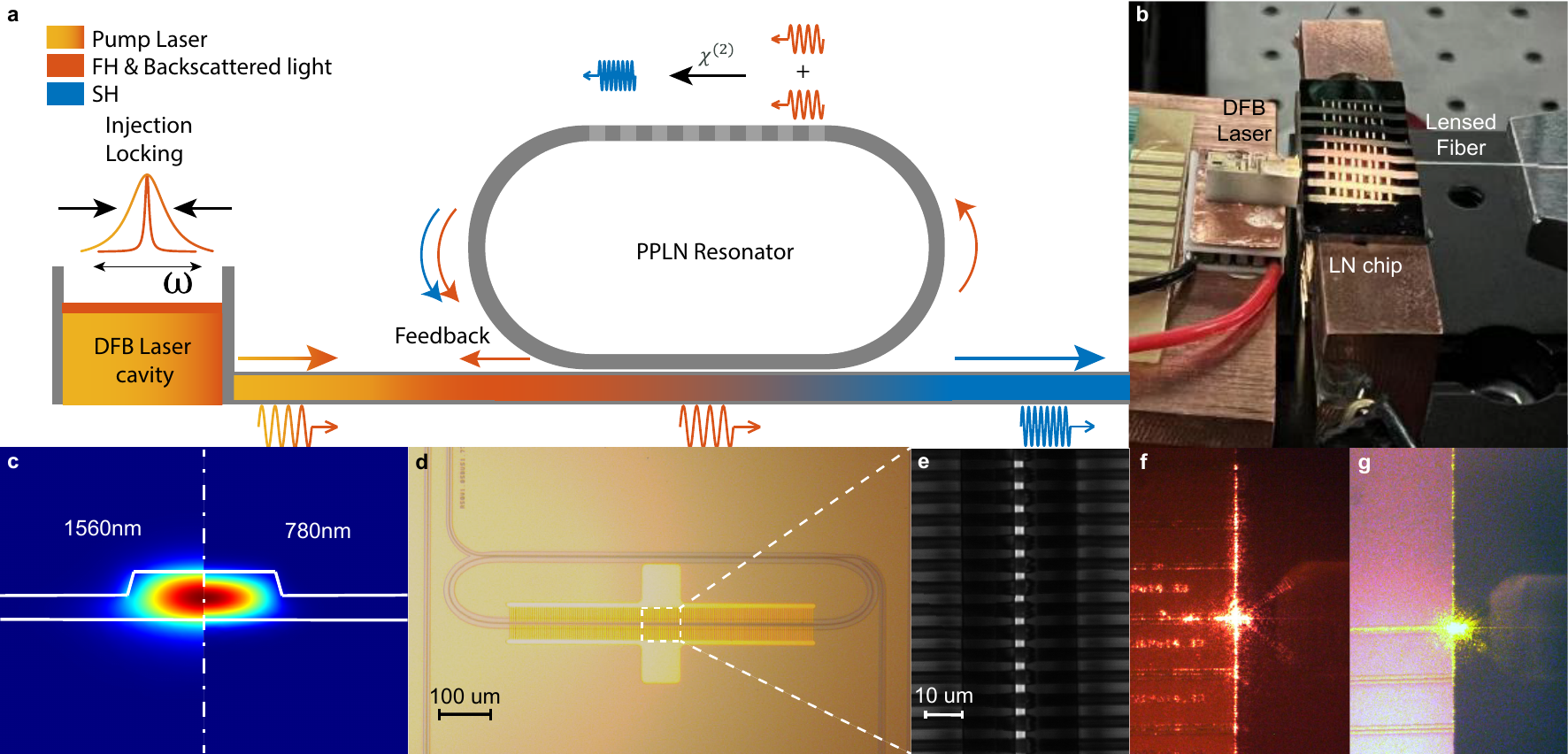}
	\caption{\label{Fig1} System concept, operating principle and resonator structure. {\bf a}, Conceptual diagram of the self-injection locking and second-harmonic generation process. {\bf b}, Optical image of the self-injection-locked frequency-doubling chip system. {\bf c}, Simulated optical mode field profile in the resonator waveguide for the fundamental quasi-TE$_{00}$ mode at 1560~nm (left) and at 780~nm (right). {\bf d}, Optical image of the PPLN racetrack resonator. Dashed box gives region of grating imaged in panel {\bf e}. {\bf e}, Second-harmonic confocal microscope image of the periodically poled waveguide section showing the uniformity of periodic poling. {\bf f} and {\bf g}, Optical images of the output facets of LN chips showing second harmonic emission generated at 780nm (red) and 589nm (yellow), respectively. }
\end{figure*}



\medskip

\noindent\textbf{Device}\nolinebreak

The high-Q lithium niobate (LN) microresonator and distributed-feedback (DFB) diode laser pump are facet-to-facet coupled as shown in Fig.~\ref{Fig1}{\bf a} and {\bf b}. The LN resonator provides both resonantly-enhanced SHG and feedback to line narrow the DFB pump. For the SHG function, it is periodically poled to quasi-phase match the doubly resonant pump and up-converted modes. For line narrowing, the high-Q mode that is pumped introduces weak backscattering into the DFB laser to achieve SIL. SIL has been shown to significantly enhance the coherence of telecom lasers \cite{liang2015ultralow,kondratiev2017self,jin2021hertz,lihachev2021ultralow,li2021reaching}, including application to soliton micro-comb generation \cite{shen2020integrated, voloshin2021dynamics,xiang2021laser}.  With the near-instantaneous nature of the SHG process, the pump coherence is readily transferred to the up-converted light, resulting in linewidth narrowing of the frequency-doubled light.


\begin{figure*}
	\centering
	\includegraphics[width=\linewidth]{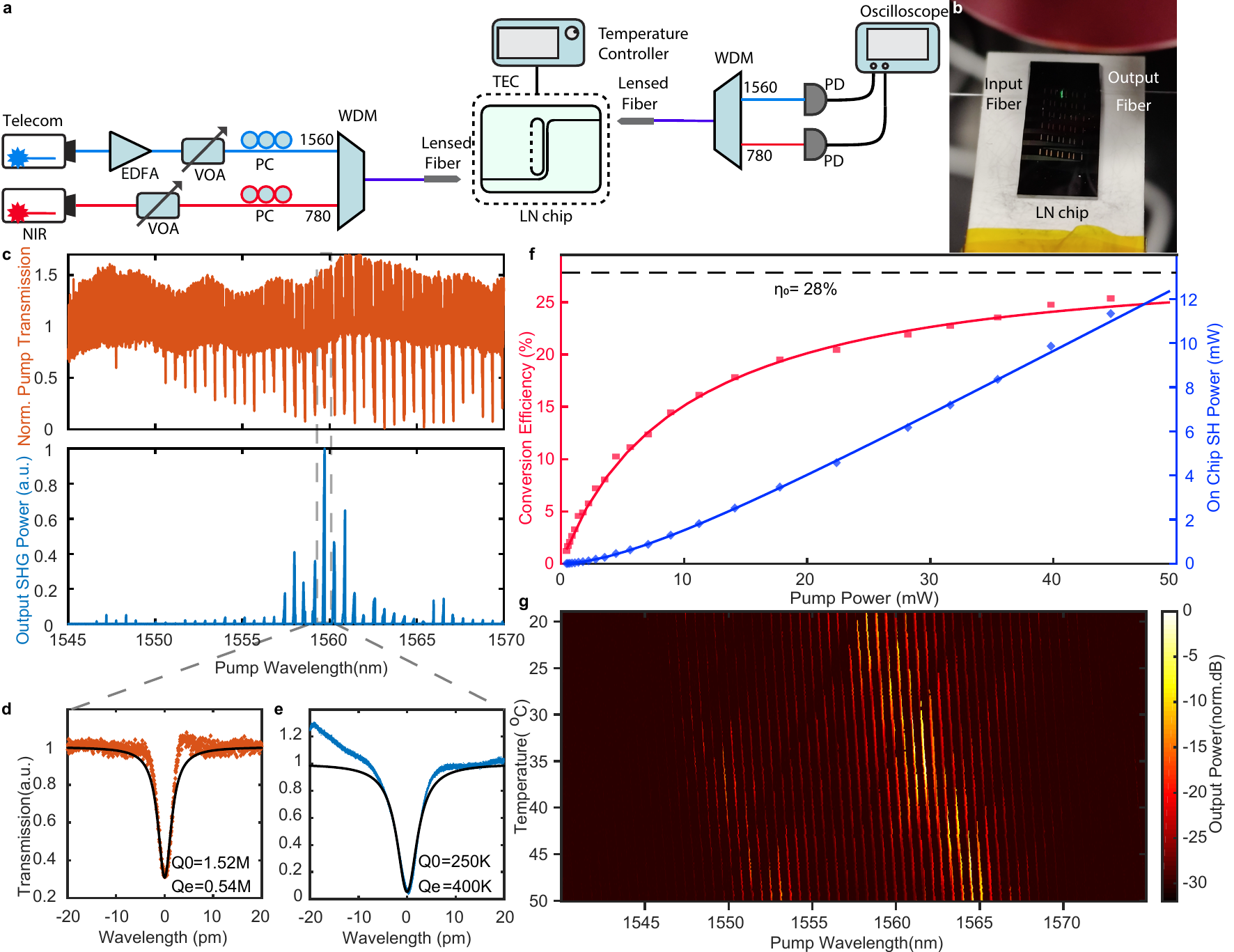}
	\caption{\label{Fig2} Characterization of the LN resonator properties and frequency-doubling performance. {\bf a}, Experimental setup to characterize the linear optical properties and the SHG performance of the resonator device. EDFA: erbium-doped fiber amplifier; VOA: variable optical attenuator; PC: polarization controller; TEC: thermoelectric cooler; WDM: wavelength division multiplexer; PD: photodiode. {\bf b}, Photograph of the LN chip showing lensed fiber coupling on both sides. {\bf c}, Transmission spectrum (upper, red) and the corresponding generated second harmonic (lower, blue) when the continous-wave laser is scanned from 1545nm to 1570nm. The device temperature is 20$^{\circ}$C. {\bf d} and {\bf e}, High resolution transmission spectrum of the pump resonance at 1559.5~nm ({\bf d}) and the SHG resonance at 779.8~nm ({\bf e}). The experimental data are shown as dots and the black solid curves show a fitting. {\bf f}, SHG power (blue) and conversion efficiency (red) as a function of the pump power on chip. The pump wavelength is fixed at 1559.5~nm. The experimental data are shown as dots and the solid curves give the theoretical fitting from a depleted SHG theory (see Supplementary Information). The dashed line indicates the theoretical maximum efficiency offered by the device. {\bf g}, Temperature dependence of the laser-scanned second-harmonic generation spectrum (also see Fig.~\ref{Fig2}{\bf c}, lower panel, blue).}
\end{figure*}

\rmfamily We utilize a race-track LN microresonator fabricated on an x-cut LN-on-insulator platform. One of the straight waveguide sections is periodically poled for quasi phase-matching of SHG, while the other section is used for external coupling (Fig.~\ref{Fig1}{\bf a}). Such a coupling approach offers a rich set of external coupling conditions through the dispersive nature of straight-waveguide coupling, and the length of the coupling region is optimized to achieve the desired coupling condition for both SIL and SHG. An Euler curve \cite{zhang2020ultrahigh} is adopted for the bending section of the racetrack resonator to suppress the bending loss so as to achieve high intrinsic optical Q. The fundamental quasi-transverse-electric modes (TE$_{00}$ modes. Fig.~\ref{Fig1}{\bf c}) are utilized for SHG to take advantage of the largest tensor element of the nonlinear susceptibility, $d_{33}$, of LN. Here we target SHG of a 1560 nm pump to a near-infrared wavelength near 780~nm where narrow-linewidth is important for atomic clock and quantum sensing applications based upon rubidium atoms \cite{ludlow2015optical, degen2017quantum}. The periodically poled section has a length of 630 $\mu$m and the resonator has a total cavity length of 1.9~mm, corresponding to a free-spectral range of 0.584 nm around the pump wavelength. 

Fig.~\ref{Fig1}{\bf d} shows a fabricated device (see Method for fabrication details). The good uniformity of the periodic poling is shown in Fig.~\ref{Fig1}{\bf e} where a second-harmonic confocal microscope image is presented. Different facet coupling approaches are utilized for the fundamental pump input and SHG light output waveguides. The pump wave is coupled onto the chip with a 5~${\rm \mu m}$-wide taper designed for laser-to-chip facet coupling. The SHG light is coupled out of the chip to a lensed fiber using a 300-nm-wide inverse tapered waveguide.  Fig.~\ref{Fig1} {\bf f} shows the 780 nm emission at the output facet with the lensed fiber also visible in the image. To demonstrate the ability to precisely engineer other SHG wavelengths, another device was developed to produce SHG at 589 nm (an important wavelength for the operation of strontium clock systems).  Yellow light emission at output facet from this device is shown in Fig.~\ref{Fig1} {\bf g}. This device was not operated in the self-injection-locking mode on account of lack of availability of a suitable DFB pumping laser.

\medskip

\noindent\sffamily\textbf{Results}\nolinebreak

\noindent\rmfamily The device is first characterized by external pumping using a tunable laser (i.e., not SIL) and the experimental setup is shown in Fig.~\ref{Fig2}{\bf a}. A telecom pump laser (Santec TSL-510) and a near-visible laser (New focus TLB-6700) characterize the linear properties of the resonator around the pump (1560~nm) and second-harmonic (780~nm) wavelengths. These are launched onto the chip via a lensed fiber (Fig.~\ref{Fig2}{\bf b}). To excite SHG, the telecom laser power is boosted by an erbium-doped fiber amplifier. The output of the chip is collected by another lensed fiber and then detected after separation of the fundamental and harmonic signals. The fiber-chip coupling loss is measured to be 6.7~dB for the pump wave and 11~dB for the second harmonic light. The device temperature is controlled by a thermo-electric cooler (TEC) to tune the phase-matching wavelength.


\begin{figure*}
	\centering
	\includegraphics[width=\linewidth]{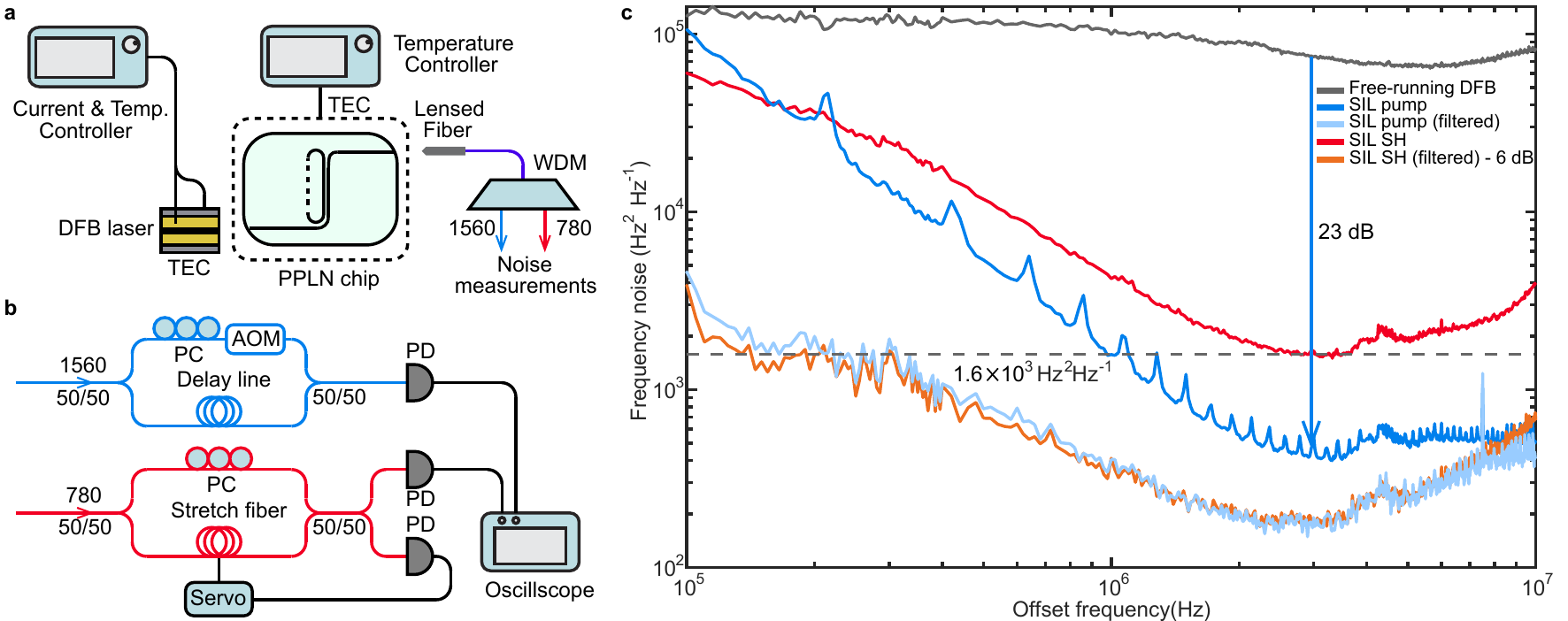}
	\caption{\label{Fig3} Self-injection-locked SHG lasing operation and experimental setup for linewidth characterization. {\bf a}, Experimental setup for SIL frequency doubling. {\bf b}  Frequency noise measurement setups for the SIL pump laser and the second harmonic light. AOM: acoustic-optical modulator. {\bf c}, Recorded frequency noise spectrum. Grey, blue and red traces show the frequency noise spectrum of the free-running DFB laser, SIL pump laser, and SIL frequency-doubled light, respectively. The light blue trace and the orange traces show the frequency noise of the pump light and 6~dB down-shifted SHG light when frequency jumping noise is removed from data.}
\end{figure*}

Fig.~\ref{Fig2}{\bf c} (upper panel) shows the transmission spectrum of the device in the telecom band. The resonator modes exhibit a range of coupling conditions from under coupled to over coupled. For pump wavelengths around $\sim$1560~nm, the device is designed to be over-coupled but with sufficient loaded Q so as to enable strong SIL while still providing relatively high SHG efficiency. As shown in Fig.~\ref{Fig2}{\bf d}, a pump resonance at 1559.5~nm exhibits an intrinsic optical Q of $Q_{1,0}={ 1.52\times10^6}$, which is slightly over-coupled with an external coupling Q of $Q_{1,e}={5.4\times10^5}$. A corresponding SHG resonance at 779.8~nm exhibits an intrinsic optical Q of $Q_{2,0}={2.5\times10^5}$ and an external coupling Q of $Q_{2,e}={4\times10^5}$, as shown in Fig.~\ref{Fig2}{\bf e}.
Fig.~\ref{Fig2}{\bf c} (lower panel) gives the SHG light generated upon spectral scan. 

Maximum conversion occurs at the pump wavelength of 1559.5~nm.  By fixing the laser wavelength at this resonance the SHG output power versus coupled power is measured in Fig.~\ref{Fig2}{\bf f}. The power of the SHG light increases quadratically with the pump power at low pump power with a slope efficiency of 3000\%/W.  At higher pump power, there is a linear dependence. Maximum SHG power is 11.3~mW at a pump power of 44.6~mW, corresponding to a conversion efficiency of about 25\%. This conversion efficiency is among the highest reported to date for on-chip LN devices \cite{wang2018ultrahigh, lu2019periodically,mckenna2021ultra,chen2021efficient}. The measurements are in good agreement with the theory (solid curves), which predicts a maximum conversion efficiency of 28\% . It was observed that optimal phase matching depended upon pumping power, through the photorefractive effect. This, in turn, required the temperature to be adjusted at each pumping power so as to maximize power conversion (see discussion below). This was also necessary in the SIL measurements discussed below. The use of magnesium-doped LN material could potentially reduce or eliminate this procedure in the future.  

The phase-matched SHG wavelength can be tuned thermo-optically as shown in Fig.~\ref{Fig2}{\bf g}.  Within a small temperature range (say, 20--25$^{o}$C), the SHG conversion maximum remains within the same resonance  and features a tuning slope of 39 pm/K that is primarily induced by the temperature dependence of refractive index.  When the device temperature increases beyond this range, the dominant SHG peak switches to a nearby mode. This is induced by the dispersion of phase matching condition, as caused by differences in the thermo-optic coefficient of LN at pump and SHG wavelengths, and by the first order dispersion of resonances.  The resulting tuning range of $\sim$7~nm over 30K  (tuning slope of about 0.23~nm/K) is useful for self-injection locking since it enables matching of the pump resonance wavelength to emission of the DFB laser.

%



For SIL frequency doubling, the setup at the input end is replaced by a DFB laser (Freedom Photonics), with direct facet-to-facet coupling to the LN chip as shown in Fig.~\ref{Fig3}{\bf a} and Fig.~\ref{Fig1} {\bf b}. The laser and the LN chip are mounted on translation stages with piezoelectric controllers. This allows precise alignment of the optical ports as well as control of the micron-scale gap between laser and chip (to set the feedback phase of the SIL process). The laser temperature is adjusted so that the output wavelength of the DFB laser roughly matches the phase-matching wavelength (1559.5~nm), and the laser current is increased to engage injection locking. At maximum pump current, the estimated on-chip pump power is 12~mW and the corresponding on-chip SHG power is 2~mW, consistent with Fig.~\ref{Fig2}{\bf f}. 

The frequency noise of the self-injection locked pump laser is characterized with a self-heterodyne approach \cite{camatel2008narrow} while the noise of SHG light is measured with a conventional homodyne detection setup with quadrature-point locking \cite{lee2012chemically} (shown in Fig.~\ref{Fig3}{\bf b}) (see Method). The results are shown in Fig.~\ref{Fig3}{\bf c}. The high-offset-frequency pump noise is significantly reduced compared to the free-running DFB laser by over 20~dB, demonstrating the effect of the SIL process. The SHG noise reaches a level of 1600~Hz$^2$~Hz$^{-1}$ at around 3~MHz offset frequency. The slight rise in this noise towards higher offset frequencies is not understood, but is not observed in another device whose test results are presented in the Supplementary Information. These data place the SHG short-term linewidth in the range of 10-30~kHz. Much larger noise suppression factors (50-60~dB) are expected if the LN resonator Q factor can be boosted to material limits reported recently \cite{gao2022lithium}.

At low offset frequencies, the frequency noise is observed to increase, and is shown in the Supplementary Information to be caused by small, random-in-time frequency jumps of the SIL laser. A further study of the jumps is presented in the Supplementary Information, including a method to isolate the jumps that enables study of their statistics as well as filtering from the frequency noise spectra. Such filtered noise spectra for the pump and the SHG signal are presented in Fig.~\ref{Fig3}{\bf c}. The origin of the frequency jumps is not currently understood.

Because the SHG process entails squaring of the pumping field, the SHG field contains exactly twice the temporal pump phase-noise fluctuation. Since the frequency noise spectrum is computed from the variance of this phase noise fluctuation, the SHG frequency noise must therefore be 4$\times$ larger (6 dB) than the pump frequency noise.  To verify this effect, the filtered SHG frequency noise data shifted by 6~dB is included in Fig.~\ref{Fig3}{\bf c}, and shows good agreement with the filtered pump frequency noise data. As the two traces are measured using independent setups with different working principles, this further validates the noise-related physics associated with the second-harmonic process. As an aside, the unfiltered SHG frequency noise in Fig.~\ref{Fig3}{\bf c} would also show closer agreement (once shifted by 6 dB) with the unfiltered pump frequency noise if not for a technical issue associated with measuring this noise in the presence of the frequency jumps. This is verified in the Supplementary Information.

\medskip

\noindent\sffamily\textbf{Discussion}\nolinebreak

\noindent\rmfamily 
The device presented here exhibits an over-coupling ratio of $\frac{Q_0}{Q_e}=2.8$ at the fundamental pump mode. Modeling shows that a similar level of over-coupling at the SHG mode can boost the conversion efficiency to above 50\%.  Also, the coupling loss in the current device is relatively high, and the linewidth of the SHG light can be further improved by reducing the laser-to-chip coupling loss using a better waveguide taper design. High coupling loss could also play a role in the observed low frequency noise behavior. Possible connection of this noise to the photo-refractive effect through mechanisms such as Barkhausen noise \cite{Barkhausen} is also under investigation. For example, switching to an MgO-doped LNOI wafer platform is being studied to significantly reduce the photorefractive effect \cite{furukawa2000photorefraction}. And any impact this has on the observed noise will be tested. 

Both the SHG process and the SIL mechanism benefit from a resonator with high intrinsic optical Q, which would allow strong over-coupling for efficient SHG while simultaneously providing a high loaded optical Q for strong SIL. Intrinsic optical Q as high as $10^8$ has recently been demonstrated in on-chip LN microresonators \cite{gao2022lithium}.  This value of optical Q is expected to further suppress the short-term linewidth of the frequency-doubled laser down to $\sim$10-Hz level.  

In summary, we have demonstrated a highly-efficient, chip-scale laser that produces high-coherence visible and near-visible light by combining self-injection locking and second-harmonic generation within a single high-Q nonlinear resonator.   A near-visible linewidth as narrow as 10~kHz is achieved by suppression of pump frequency noise. Also, on-chip converted power over 2 mW is obtained in a hybrid-integrated design. Separate measurements using an external pump laser yield a maximum conversion efficiency over 25\% and maximum SHG power of 11.3~mW. These measurements suggest that  much higher integrated device performance will be possible by replacing the current hybrid design with a heterogeneously-integrated device that features low pump to LN resonator coupling loss.  
The demonstrated approach can be applied to other optical frequency conversion processes such as optical parametric oscillation for frequency down-conversion, and third-harmonic generation for up-conversion to shorter wavelengths. Moreover, the on-chip LN platform enables integration with electro-optic components for further functional enhancement. Such a highly coherent, chip-scale visible/near-visible optical source can profoundly impact many applications including optical clocks, metrology, sensing, and quantum information processing.


\medskip

\noindent\sffamily\textbf{Method}\nolinebreak

\noindent\rmfamily\textbf{Device fabrication}
The device is fabricated on a congruent x-cut thin film lithium-niobate-on-insulator (LNOI) wafer, with 600~nm LN sitting upon a 4.7~$\rm \mu m$ silica layer. ZEP-520A resist is used as the mask for a first e-beam lithography step (EBL), followed by 300~nm Ar-ion milling to define the waveguide. Next, a second EBL writing step is performed on PMMA resist, and 400~nm electrodes are created using a gold evaporation and lift-off process.  A 3 $\mu$m separation is introduced between the electrodes and the waveguide to reduce the absorption. Finally, the chip is diced and polished to acquire good fiber-to-chip and laser-to-chip coupling.
\\\indent The periodic poling process is similar to the previously reported methods\cite{zhao2020shallow}, but the poling is done after the etching process. Several pulses of 10 ms duration and voltage of 240 V are applied with the electric field along the negative z axis of the LN crystal.  The SHG power is monitored during poling until SHG power is maximized. The number of pulses used is around 20, and the poling uniformity can be clearly observed in Fig.~\ref{Fig2}{\bf b}. A poling period of 4.3$\mu$m is used to compensate for the momentum mismatch between the pump and SHG in the 1.75$\mu$m waveguide.\\

\medskip

\noindent\rmfamily\textbf{DFB Laser}
The pump laser used in this work is an InP distributed feedback laser (DFB) as shown in Fig.~\ref{Fig3}{\bf d}. It has high-reflection coatings on the back facet and anti-reflection coatings on the front facet. The threshold current is ~50 mA, and the total output power of the laser can reach more than 100~mW under 400 mA pumping current. The slope efficiency of the laser is around 0.28~$\rm mW\cdot mA^{-1}$ and the peak wall-plug efficiency can reach 20\%. The lasing is single-mode and its wavelength also shifts from  ~1,558.2~nm to ~1,560.2~nm with increasing bias current (about -0.75 $\rm GHz\cdot mA^{-1}$). The laser chip is mounted on a sub-mount with a thermoelectric cooler underneath.  The temperature of the laser diode can be controlled within 1 mK.\\

\medskip

\noindent\rmfamily\textbf{Noise measurements}
The self-injection-locked laser pump at 1559.5~nm is characterized by a self-heterodyne method (see Fig.~\ref{Fig3}{\bf b} upper branch in blue) . A portion of the laser is frequency-downshifted by an AOM and recombined with another portion delayed by a 1-km-long fiber. The optical signal is received by a balanced photodetector, converted to a radio-frequency signal, and recorded with a high-speed oscilloscope. Phase fluctuations are extracted using Hilbert transforms and then converted to frequency fluctuation through time-domain difference. Other details of the setup are documented elsewhere \cite{wang2020towards}.

The second harmonic light at 779.8~nm is characterized by a homodyne method  (see Fig.~\ref{Fig3}{\bf b} lower branch in red). The homodyne method (as opposed to the heterodyne method describe above) was used at 779.8~nm, because existing AOM optical losses at this wavelength were too large. The system is stabilized around the quadrature point using a servo loop to control a stretched optical fiber, such that phase fluctuation can be linearly converted to voltage.  Compared to the self-heterodyne method, the laser is not frequency-shifted, and the stretch fiber is equivalent to a 15 m delay. The output signal is sent to the same oscilloscope for data recording, which is synchronized with the pump laser signal for comparison. Phase fluctuations can be calculated by normalizing the signal against the voltage swing, measured separately by applying a sweeping signal at the stretch fiber. Other details of the calculation are documented elsewhere \cite{lee2012chemically}. 

\medskip

\noindent\rmfamily\textbf{Data availability.} The data that support the findings of this study are available from the corresponding author upon reasonable request. 

\bibliographystyle{naturemag} 
\bibliography{ref.bib}

	\medskip

\noindent\sffamily\textbf{Acknowledgements}

\noindent\rmfamily 	This work is supported in part by the Defense Advanced Research Projects Agency (DARPA) LUMOS program under Agreement No.~HR001-20-2-0044, the Defense Threat Reduction Agency-Joint Science and Technology Office for Chemical and Biological Defense (grant No.~HDTRA11810047), and the National Science Foundation (NSF) (ECCS-1810169, ECCS-1842691 and, OMA-2138174). This work was performed in part at the Cornell NanoScale Facility, a member of the National Nanotechnology Coordinated Infrastructure (National Science Foundation, ECCS-1542081); and at the Cornell Center for Materials Research (National Science Foundation, Grant No. DMR-1719875). \newline

\medskip

\noindent\sffamily\textbf{Author Contributions}\nolinebreak

\noindent\rmfamily J.L., J.S., B.Q., H.W., L.C., J.B., K.V. and Q.L. conceived the experiment. J.L., J.S designed and fabricated the PPLN device and L.C. designed and fabricated the DFB laser. J.L., J.S., B.Q. and H.W. carried out the device characterization. U.J., L.W. and M.L. assisted in the device fabrication. U.J., R.L., Y.H. and Z.Y. assisted in experiments.  J.L., J.S., B.Q., H.W., L.C. and Q.L. wrote the manuscript with contribution from all authors. Q.L., K.V. and J.B. supervised the project.\newline

\noindent\sffamily\textbf{Additional Informations}

\noindent\rmfamily\textbf{Reprints and permission.} Reprints and permissions are available at www.nature.com/reprints.  

\noindent\textbf{Competing interests.} The authors declare no competing interests.  Correspondence and requests for materials should be sent to  Q.L. (qiang.lin@rochester.edu), K.V. (vahala@caltech.edu).), or J.B. (jbowers@ucsb.edu).

\medskip
\if{
\noindent\textbf{Paragraphs moved from the Main text. Need to be placed into the Supplement.} 

\textcolor{red}{This paragraph should go later or even in the Supplement.}
For resonantly enhanced SHG in a high-Q resonator, the maximum conversion efficiency and the required pump power are given by the following expressions (see Supplementary Information)
\begin{equation}\label{maxes}
\eta_0=\frac{\kappa_{1e}\kappa_{2e}}{\kappa_{1t}\kappa_{2t}},\qquad 
P_{0}=\hbar\omega_1 \frac{\kappa_{1t}^3\kappa_{2t}}{8\kappa_{1e}|\gamma|^2}
\end{equation}
where $\kappa_{je}$ and $\kappa_{jt}$ are the external coupling rate and photon decay rate of the loaded cavity, for the pump ($j=1$) and second-harmonic mode ($j=2$), respectively. $\kappa_{jt} = \kappa_{je} + \kappa_{j0}$ where $\kappa_{j0}$ is the photon decay rate of the intrinsic cavity. $\gamma$ is the nonlinear coupling rate for the SHG process. Equation (\ref{maxes}) shows that high conversion efficiency requires strong over coupling at both the pump and second harmonic resonances ($\kappa_{je} > \kappa_{j0}$), as shown in Fig.~\ref{Fig1}{\bf c}. This is because SHG coexists with the parametric down-conversion process in the resonator (i.e.,the conversion is reversible such that $\omega_1 + \omega_1 \leftrightarrow \omega_2$). As a result, high SHG efficiency requires the second harmonic light to be quickly extracted from the cavity to prevent its conversion back into the fundamental frequency. Eq.~(\ref{maxes}) also shows that the required pump power depends strongly on the loaded cavity Q of the resonator ($Q = \omega/\kappa$). High optical Q, particularly at the pump resonance, is crucial for reducing the pump power. So far, most research efforts \cite{chang2016thin, wolf2018quasi, wang2018ultrahigh, luo2019optical, lin2019broadband, rao2019actively, chen2019ultra,  lu2019periodically, niu2020optimizing, zhao2020shallow, lu2020toward, mckenna2021ultra, chen2021efficient, wang2021efficient} have been focused on achieving a low pump power required for SHG or a high slope efficiency for frequency conversion (which is related to $\gamma$ as $\rho = 64 |\gamma|^2 \kappa_{2e} \kappa_{1e}^2/(\kappa_{2t}^2 \kappa_{1t}^4)$). However, in practice, the converted power (given by $P_0 \eta_0 = \frac{\hbar \omega}{8 |\gamma|^2} \kappa_{2e} \kappa_{1t}^2$) at the point of maximum conversion efficiency is important. In these cases where converted power is of major concern, a high loaded optical Q might not be favorable. 

\indent \textcolor{red}{This paragraph should go later or even in the Supplement.} The situation is different for SIL because strength depends on the amplitude of the pump wave that is scattered back into the DFB laser. The enhancement of the laser coherence is characterized by the noise reduction factor defined as the ratio between the noise spectral intensity of the original DFB laser, $S^{(f)}_{\rm DFB}$, and that of the SIL state, $S^{(f)}_{\rm SIL}$, $F_{\rm NR} \equiv \frac{S^{(f)}_{\rm DFB}}{S^{(f)}_{\rm SIL}}$. For the SIL frequency doubling system shown in Fig.~\ref{Fig1}{\bf a}, it is given by (see Supplementary Information) 
\begin{multline}\label{NRF}
F_{\rm NR} \approx \\64(1+\alpha^2)|\beta|^2\frac{Q_{1t}^2}{Q_{\rm DFB}^2}\frac{\kappa_{1e}^2}{\kappa_{1t}^2}T^2\left(\frac{1+(1/2-\kappa_{1t}/\kappa_{2t})C}{(1+C)^2}\right)^2
\end{multline}
where $\alpha$ is the amplitude-phase coupling factor of the DFB laser, $\beta$ is the back-scattering factor of the LN resonator, $T$ is the facet transmission between the two chips, $C$ is the cooperativity of the SHG process, $Q_{1t}$ and $Q_{\rm DFB}$ are the loaded Q of the LN resonator and the DFB laser, respectively, at the pump frequency. As shown in Eq.~(\ref{NRF}), the noise reduction factor is sensitive to the optical Q of the pump resonance, while it only has a minor dependence on that of the SH mode. Fig.~\ref{Fig1}{\bf d} shows this feature. Stong SIL requires a high loaded optical Q at the pump resonance. The primary reason lies in the fact that the phase dependence of the pump on the back-scattered light from a high-Q resonator relies on the locking strength which is proportional to$|\beta| Q_{1t}$. The higher optical Q, the larger phase response of the back-scattered pump wave and thus the stronger SIL. 

\textcolor{red}{This paragraph should go later or even in the Supplement.} Apparently, SHG and SIL have different requirements on the device, the former favors strong over-coupling while the latter requires high loaded optical Q at the pump frequency. As the intrinsic optical Q of the SHG resonator is primarily determined by the fabrication quality in practice, over-coupling the device would increase the maximum conversion efficiency while decreasing the loaded Q of LN resonator and the strength of SIL. Efficient SIL-frequency doubling would thus require an appropriate balance between these two processes.  
}\fi
\end{document}



\title{Supplementary Information for\\{\bf "Self-injection locked frequency doubling" }}
\author{Jingwei Ling}
\thanks{These authors contributed equally.}
\affiliation{Department of Electrical and Computer Engineering, University of Rochester, Rochester, NY 14627, USA}
\author{Jeremy Staffa}
\thanks{These authors contributed equally.}
\affiliation{Institute of Optics, University of Rochester, Rochester, NY 14627, USA}
\author{Boqiang Shen}
\thanks{These authors contributed equally.}
\affiliation{T. J. Watson Laboratory of Applied Physics, California Institute of Technology, Pasadena, CA 91125, USA}
\author{Heming Wang}
\thanks{These authors contributed equally.}
\affiliation{T. J. Watson Laboratory of Applied Physics, California Institute of Technology, Pasadena, CA 91125, USA}
\author{Lin Chang}
\thanks{These authors contributed equally.}
\affiliation{Department of Electrical and Computer Engineering, University of California Santa Barbara, Santa Barbara, CA 93106, USA}
\author{Usman A. Javid}
\affiliation{Institute of Optics, University of Rochester, Rochester, NY 14627, USA}
\author{Lue Wu}
\affiliation{T. J. Watson Laboratory of Applied Physics, California Institute of Technology, Pasadena, CA 91125, USA}
\author{Zhiquan Yuan}
\affiliation{T. J. Watson Laboratory of Applied Physics, California Institute of Technology, Pasadena, CA 91125, USA}
\author{Raymond Lopez-Rios}
\affiliation{Institute of Optics, University of Rochester, Rochester, NY 14627, USA}
\author{Mingxiao Li}
\affiliation{Department of Electrical and Computer Engineering, University of Rochester, Rochester, NY 14627, USA}
\author{Yang He}
\affiliation{Department of Electrical and Computer Engineering, University of Rochester, Rochester, NY 14627, USA}
\author{Bohan Li}
\affiliation{T. J. Watson Laboratory of Applied Physics, California Institute of Technology, Pasadena, CA 91125, USA}
\author{John E. Bowers}
\email{jbowers@ucsb.edu}
\affiliation{Department of Electrical and Computer Engineering, University of California Santa Barbara, Santa Barbara, CA 93106, USA}
\author{Kerry J. Vahala}
\email{vahala@caltech.edu}
\affiliation{T. J. Watson Laboratory of Applied Physics, California Institute of Technology, Pasadena, CA 91125, USA}
\author{Qiang Lin}
\email{qiang.lin@rochester.edu}
\affiliation{Department of Electrical and Computer Engineering, University of Rochester, Rochester, NY 14627, USA}
\affiliation{Institute of Optics, University of Rochester, Rochester, NY 14627, USA}



\begin{abstract}
In this supplement detailed information is provided on the following: the couple mode theory for second harmonic generation; the theory for self-injection locking with focus placed on the noise reduction in the SHG process.
\end{abstract}
\maketitle
\section{Theory}
\subsection{Design for LN resonator}
\begin{figure}
    \centering
	\includegraphics[width=\linewidth]{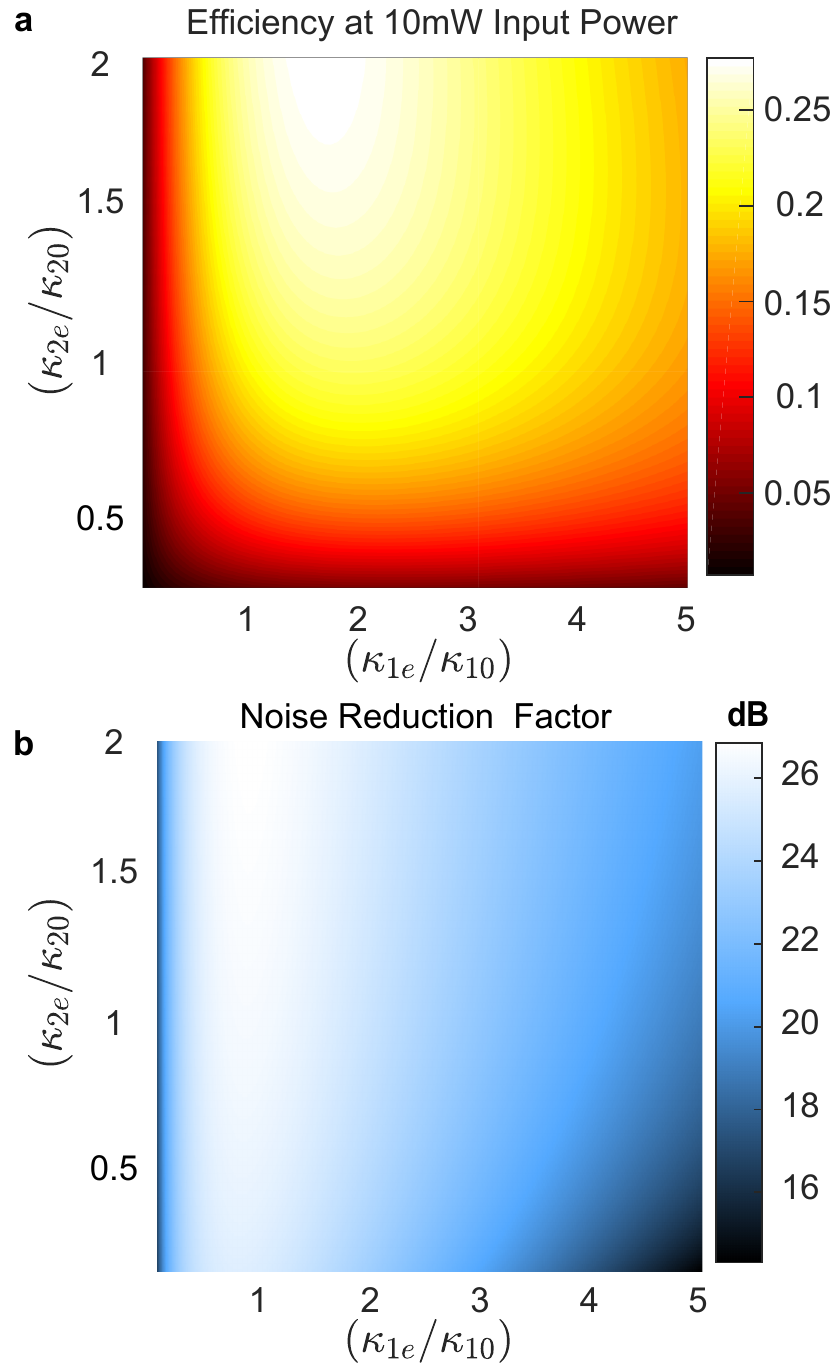}
	\caption{\label{FigS1} Design principles.  {\bf a},{\bf b}, Numerically simulated conversion efficiency {\bf a}, and noise reduction factor {\bf b}, at different coupling conditions, assuming fixed intrinsic Q, and other conditions optimized for maximum conversion efficiency.}
\end{figure}

For resonantly enhanced SHG in a high-Q resonator, the maximum conversion efficiency and the required pump power are given by the following expressions (see next subsection for details)
\begin{equation}\label{efficiency}
\eta_0=\frac{\kappa_{1e}\kappa_{2e}}{\kappa_{1t}\kappa_{2t}},\qquad 
P_{0}=\hbar\omega_1 \frac{\kappa_{1t}^3\kappa_{2t}}{8\kappa_{1e}|\gamma|^2}
\end{equation}
where $\kappa_{je}$ and $\kappa_{jt}$ are the external coupling rate and photon decay rate of the loaded cavity, for the pump ($j=1$) and second-harmonic mode ($j=2$), respectively. $\kappa_{jt} = \kappa_{je} + \kappa_{j0}$ where $\kappa_{j0}$ is the photon decay rate of the intrinsic cavity. $\gamma$ is the nonlinear coupling rate for the SHG process. Equation (\ref{efficiency}) shows that high conversion efficiency requires strong over coupling at both the pump and second harmonic resonances ($\kappa_{je} > \kappa_{j0}$), as shown in Fig.~\ref{FigS1}{\bf a}. This is because SHG coexists with the parametric down-conversion process in the resonator (i.e., the conversion is reversible such that $\omega_1 + \omega_1 \leftrightarrow \omega_2$). As a result, high SHG efficiency requires the second harmonic light to be quickly extracted from the cavity to prevent its conversion back into the fundamental frequency. Eq.~(\ref{efficiency}) also shows that the required pump power depends strongly on the total loss, or, in other word, the loaded cavity Q of the resonator ($Q = \omega/\kappa$). High optical Q, particularly at the pump resonance, is crucial for reducing the pump power. So far, most research efforts \cite{chang2016thin, wolf2018quasi, wang2018ultrahigh, luo2019optical, lin2019broadband, rao2019actively, chen2019ultra,  lu2019periodically, niu2020optimizing, zhao2020shallow, lu2020toward, mckenna2021ultra, chen2021efficient, wang2021efficient} have been focused on achieving a low pump power required for SHG or a high slope efficiency for frequency conversion (which is related to $\gamma$ as $\rho = 128 |\gamma|^2 \kappa_{2e} \kappa_{1e}^2/(\hbar\omega_1\kappa_{2t}^2 \kappa_{1t}^4)$). However, in practice, the converted power (given by $P_0 \eta_0 = \frac{\hbar \omega}{8 |\gamma|^2} \kappa_{2e} \kappa_{1t}^2$) at the point of maximum conversion efficiency is important. In these cases where converted power is of major concern, a high loaded optical Q might not be favorable. 

Concerning the SIL process, the enhancement of the laser coherence is characterized by the noise reduction factor defined as the ratio between the noise spectral intensity of the original DFB laser, $S^{(f)}_{\rm DFB}$, and that of the SIL state, $S^{(f)}_{\rm SIL}$, $F_{\rm NR} \equiv \frac{S^{(f)}_{\rm DFB}}{S^{(f)}_{\rm SIL}}$, which is given by (see next subsection for derivation): 
\begin{multline}\label{NRF}
F_{\rm NR} \approx \\64(1+\alpha^2)|\beta|^2\frac{Q_{1t}^2}{Q_{\rm DFB}^2}\frac{\kappa_{1e}^2}{\kappa_{1t}^2}T^2\left(\frac{1+(1/2-\kappa_{1t}/\kappa_{2t})C}{(1+C)^2}\right)^2
\end{multline}
where $\alpha$ is the amplitude-phase coupling factor of the DFB laser, $\beta$ is the back-scattering factor of the LN resonator, $T$ is the facet transmission between the two chips, $C$ is the cooperativity of the SHG process, $Q_{1t}$ and $Q_{\rm DFB}$ are the loaded Q of the LN resonator and the DFB laser, respectively, at the pump frequency. As shown in Eq.~(\ref{NRF}), the noise reduction factor is sensitive to the optical Q of the pump resonance, while it only has a minor dependence on that of the SH mode. Fig.~\ref{FigS1}{\bf b} shows this feature. Strong SIL requires a high loaded optical Q at the pump resonance. The primary reason lies in the fact that the phase dependence of the pump on the back-scattered light from a high-Q resonator relies on the locking strength which is proportional to$|\beta| Q_{1t}$. The higher the optical Q, the larger the phase response of the back-scattered pump wave and thus the stronger SIL. 

From the above discussion, SHG and SIL impose different requirements on the device. The former favors strong over-coupling while the latter requires high loaded optical Q at the pump frequency. As the intrinsic optical Q of the SHG resonator is primarily determined by the fabrication quality in practice with the coupling Q adjustable, over-coupling the device would increase the maximum conversion efficiency while decreasing the loaded Q of LN resonator and the strength of SIL. Efficient SIL-frequency doubling thus requires an appropriate balance between these two processes.  
\subsection{Coupled Mode Theory of Resonant SHG}
The intra-cavity SHG process is described by a pair of coupled mode equations for the cavity photon number amplitudes \cite{luo2019optical}:
\begin{align}\label{SHGCMT}
\frac{\mathrm{d}a_1}{\mathrm{d}t}=&\left(i\delta_1-{\kappa_{1t}}/{2}\right)a_1+2i\gamma^\ast a_1^\ast a_2+i\sqrt{\kappa_{1e}}f_1\\
\frac{\mathrm{d}a_2}{\mathrm{d}t}=&\left(i\delta_2-{\kappa_{2t}}/{2}\right)a_2+i\gamma a_1^2
\end{align}
where in $\delta_1=\omega_1-\omega_{10}$ is the pump-cavity detuning, $\delta_2=2\delta_1-\Delta\omega$ is the detuning from the SH resonance, $\Delta\omega=\omega_{20}-2\omega_{10}$ is the mismatch between the fundamental (FH) and second-harmonic (SH) cavity resonances, $\omega_1$ is the angular FH pumping frequency, $\omega_{10}$ and $\omega_{20}$ is the angular resonance frequency for the FH and SH modes, respectively, $\kappa_{jt}=\kappa_{j0}+\kappa_{je}$ are the total linewidths of the resonances comprised of the intrinsic and external loss rates, $\gamma$ is the nonlinear coupling rate, and $f_1$ is the input photon flux amplitude from the pump such that $P_{\text{in}}=\hbar\omega_1|f_1|^2$, and the output power flux can be written as  $P_{\text{out}}=\hbar\omega_2\kappa_{2e}|a_2|^2$.
In steady state, the coupled equations can be solved as
\begin{align}
|a_1|^2=&\;4\frac{\kappa_{1e}|f_1|^2}{\kappa_{1t}^2(1+\ell_1^2)}\left|1+\frac{C}{(1-i\ell_1)(1-i\ell_2)}\right|^{-2}\label{a1sol}\\
|a_2|^2=&\;4\frac{|\gamma|^2|a_1|^4}{\kappa_{2t}^2(1+\ell_2^2)}\label{a2sol}
\end{align}
where we have defined the linewidth-normalized detunings $\ell_j=\frac{2\delta_j}{\kappa_{jt}}$ and the resonant cooperativity factor $C=\frac{8|\gamma|^2\mid a_1 \mid^2}{\kappa_{1t}\kappa_{2t}}$.  The total conversion efficiency in general can be expressed as the ratio between the output SH power flux and the input pump power flux \begin{align}
\eta=\frac{P_{out}}{P_{in}}=\eta_0\frac{\frac{4C}{(1+\ell_1^2)(1+\ell_2^2)}}{\left|1+\frac{C}{(1-i\ell_1)(1-i\ell_2)}\right|^2}
\end{align}
and, in perfect resonant conditions $\ell_1=\ell_2=0$, this yields,
\begin{align}
\eta=\eta_0\frac{4C}{(1+C)^2}
\end{align}
where the maximum efficiency $\eta_0$ is achieved when $C=1$ corresponding to input power $P_0$, as written in Eq.~(\ref{efficiency}).
For imperfect resonant conditions, the maximum efficiency condition becomes $C=\sqrt{1+\ell_1^2}\sqrt{1+\ell_2^2}$, and achieves reduced maximum efficiency $\eta_0^{'}$ at increased input power $P_0^{'}$, each expressed as
\begin{align}
\eta_0^{'}=&\eta_0\frac{2}{\sqrt{(1+\ell_1^2)(1+\ell_2^2)}+(1-\ell_1\ell_2)}\\
P_{0}^{'}=&P_0\frac{\sqrt{1+\ell_1^2}}{2}\left(1+\ell_1\ell_2+\sqrt{(1+\ell_1^2)(1+\ell_2^2)} \right)
\end{align}
It is worth noting that, in general, the cooperativity $C$ satisfies the equation
\begin{multline}
4\frac{P_{\text{in}}}{P_0}=(1+\ell_1^2)\Bigg(C+2C^2\frac{1-\ell_1\ell_2}{(1+\ell_1^2)(1+\ell_2^2)}\\+C^3  \frac{1}{(1+\ell_1^2)(1+\ell_2^2)}\Bigg)
\end{multline}
and, for perfect resonant conditions, the simplified equation
\begin{align}
\label{cubic solution}
4\frac{P_{\text{in}}}{P_0}=C+2C^2+C^3
\end{align}
thereby connecting the cooperativity $C$ with the input power $P_{\text{in}}$. These are cubic polynomials in $C$, which obviously have well known analytic solutions. From our definitions, a solution for $C$ is equivalent to a solution for the intra-cavity photon number $|a_1|^2$ and consequently $|a_2|^2$ as well. One thing to notice is that, in the testing section of the maintext, the $P_0=151 mW$ and $\eta_0=28\%$ require high input and are not attained. By solving the Eq.~(\ref{cubic solution}) for the experimental input, we get an estimated efficiency curve matching well with the experimental result, with highest efficiency of 25\% reached at 44.6 mW input.

\subsection{Theory for self-injection locking (SIL) process}
In addition to the above equations for $a_1$ and $a_2$, two additional equations are introduced to describe the SIL process. The first one is a field equation for the photon number amplitude $b_1$ as pumped by the backscattered mode:
\begin{equation}
\frac{\mathrm{d}b_1}{\mathrm{d}t}=\left(i\delta_1-\frac{\kappa_{1t}}{2}\right)b_1+i\beta\frac{\kappa_{1t}}{2}a_1\\
\end{equation}
where $\beta$ is the normalized backscattering coefficient for the FH mode. We assume that $\beta \ll 1$ such that $|b_1|^2$ is small compared to $|a_1|^2$ and neglect the SHG process associated with $b_1$. We also ignore the scattering process from $b_1$ back to $a_1$. The small $\beta$ assumption is justified by the fact that no mode splitting has been observed for the fundamental mode.

The second equation is an algebraic relation for the pumping frequency \cite{kondratiev2017self,shen2020integrated}:
\begin{equation}\label{locking}
\omega_1 = \omega_{\text{L}} - \frac{\kappa_{1t}}{2}K\text{Im}\left[\mathrm{e}^{i\phi}\frac{\kappa_{1t}}{2}\frac{b_1}{(i\beta)(i\sqrt{\kappa_{1e}} f_1)}\right]
\end{equation}
where $\omega_{\text{L}}$ is the free-running frequency of the DFB laser, $\phi$ is the round-trip feedback phase, $K$ is the normalized locking strength:
\begin{equation}
K=4\frac{Q_{1t}}{Q_{\rm DFB}}\frac{\kappa_{1e}}{\kappa_{1t}}|\beta|T\sqrt{1+\alpha^2}
\end{equation}
$\alpha$ is the amplitude-phase coupling factor of the DFB laser,  and $T$ is the single-trip transmission between laser and resonator (e.g. loss at the coupling facet). In arriving at this equation, we have assumed that the laser gain saturation is strong enough such that the output power remains approximately constant, and the laser carrier dynamics are much faster than the resonator relaxation time (as determined by $\kappa_{1t}^{-1}$) such that the frequency adiabatically follows the injection-locked equilibrium. Within these assumptions, the injection locking process becomes independent of the DFB laser gain details, which simplifies the following analysis. We will also focus on perfect phase-matching conditions for the SHG process by setting $\ell_1=\ell_2=0$ at steady state.

To study the noise reduction effect, we assume that a slight change in $\omega_{\text{L}}$ induces a change in $\omega_1$, the detunings and all field amplitudes. Using $\omega_1$ as the dependent variable, The change of $|a_1|^2$ can be written as, using Eq. (\ref{a1sol}) at $\ell_1=\ell_2=0$,
\begin{equation}
\frac{\partial|a_1|^2}{\partial\omega_1}=4\frac{\kappa_{1e}|f_1|^2}{\kappa_{1t}^2}\frac{2}{(1+C)^3}\frac{\partial C}{\partial\omega_1}
\end{equation}
However, since the cooperativity factor is directly proportional to $|a_1|^2$, we conclude that $\partial|a_1|^2/\partial\omega_1=0$ at perfect phase-matching. Similar arguments apply to $a_2$ and $b_1$, where there are no changes of field amplitudes up to first order of $\omega_1$ at perfect phase-matching.

The steady-state $a_1$ can be solved from Eq. (\ref{SHGCMT}) and reads
\begin{equation}
a_1=\frac{2i\sqrt{\kappa_{1e}}f_1}{\kappa_{1t}(1-i\ell_1)}\left(1+\frac{C}{(1-i\ell_1)(1-i\ell_2)}\right)^{-1}
\end{equation}
and its derivative reads
\begin{equation}
\frac{\partial{a_1}}{\partial\omega_1}=\frac{2\sqrt{\kappa_{1e}}f_1}{\kappa_{1t}}\frac{1}{(1+C)^2}\left(\frac{4C}{\kappa_{2t}}-\frac{2}{\kappa_{1t}}\right)
\end{equation}
where we have used $\partial\ell_1/\partial\omega_1=2/\kappa_{1t}$ and $\partial\ell_2/\partial\omega_1=4/\kappa_{2t}$. Similarly for $b_1$ we have
\begin{equation}
b_1=\frac{i\beta}{1-i\ell_1}a_1
\end{equation}

\begin{multline}
\frac{\partial b_1}{\partial \omega_1}=\frac{2i\beta(i\sqrt{\kappa_{1e}}f_1)}{\kappa_{1t}}\\\times\Bigg(i\frac{2/\kappa_{1t}}{1+C}-i\frac{4C/\kappa_{2t}-2/\kappa_{1t}}{(1+C)^2}\Bigg)
\end{multline}

Finally, from the locking relation Eq. (\ref{locking}),
\begin{equation}
\frac{\partial \omega_{\text{L}}}{\partial \omega_1}=1+2K\cos\phi\frac{1+(1/2-\kappa_{1t}/\kappa_{2t})C}{(1+C)^2}
\end{equation}
The first term can be dropped if $K \gg 1$. For optimal locking that maximizes $\partial \omega_{\text{L}}/\partial \omega_1$, the feedback phase should be chosen as $\phi=0$ (the other choice $\phi = \pi$ leads to an unstable equilibrium). Replacing $K$ with measurable quantities, the noise reduction factor can be expressed as
\begin{multline}
F_{\rm NR}  = \Bigg|\frac{\partial \omega_{\text{L}}}{\partial \omega_1}\Bigg|^2
 \approx 64(1+\alpha^2)|\beta|^2\frac{Q_{1t}^2}{Q_{\rm DFB}^2}\frac{\kappa_{1e}^2}{\kappa_{1t}^2}T^2\\\times\left(\frac{1+(1/2-\kappa_{1t}/\kappa_{2t})C}{(1+C)^2}\right)^2
\end{multline}
which corresponds to Eq. (\ref{NRF}) in the first subsection.

\section{Discussion of noise measurements}

\begin{figure}
    \centering
	\includegraphics[width=\linewidth]{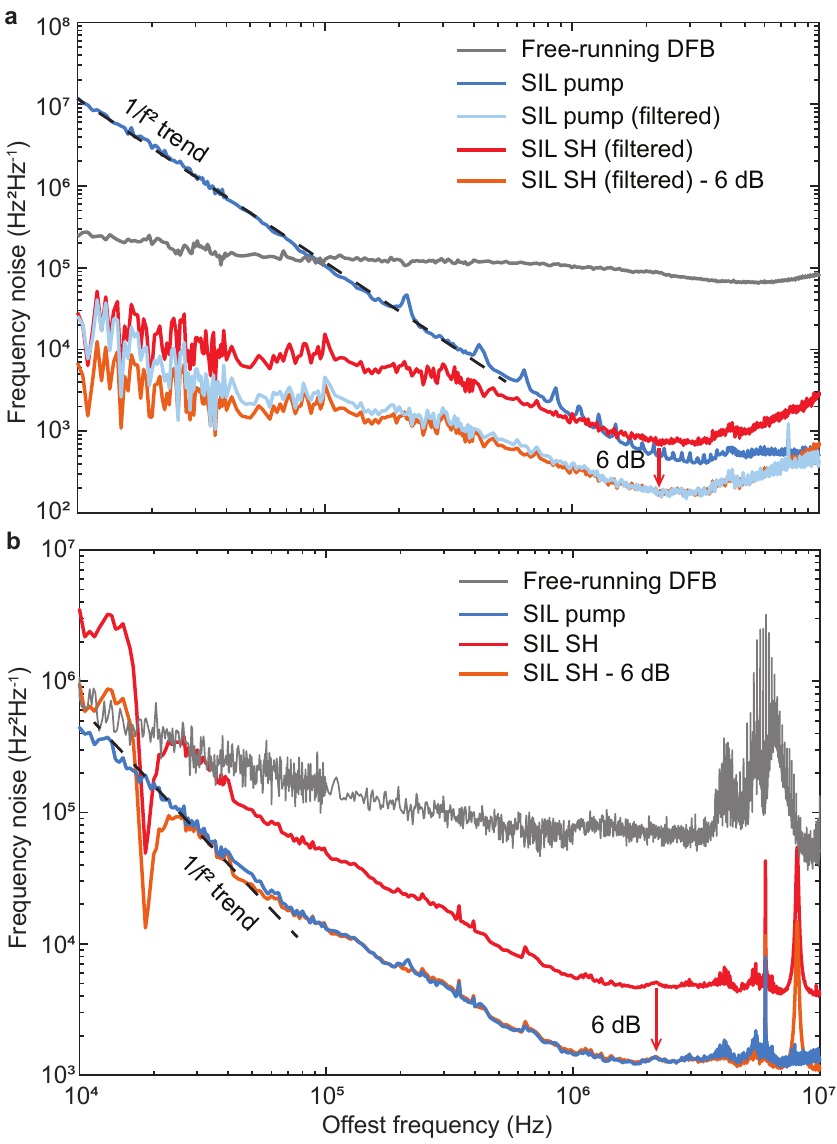}
	\caption{\label{FigS2} 
	Frequency noise measurement using different DFB pump lasers and LN devices.
	{\bf a}, Frequency noise measurement starting at a lower offset frequency versus that in Main text. Black and light blue traces show the frequency noise spectrum of the free-running DFB laser and SIL pump laser. The short term noise level is suppressed by about 20 dB resulting in a high offset frequency noise of about 400 Hz$^2$/Hz (equivalent short term linewidth of 2.5 kHz and second harmonic linewidth of 10 kHz). Frequency spectra obtained for the pump and second-harmonic signals after filtering are shown as blue and red traces for comparison. Also, the theoretical noise spectrum ($f^{-2}$) produced by Poissonian frequency noise is plotted for comparison (dashed line).
	{\bf b}, Frequency noise measurement based on a different DFB pump laser and a different LN device relative to Main text. Grey, blue, red lines show the frequency noise spectrum of the free-running DFB laser, SIL pump light and SIL SHG light. The orange curve shows the 6~dB down shifted noise curve for SHG, agreeing well with the pump light. Filtering for frequency steps is not used in this measurement. The short-term noise level is suppressed by about 17 dB resulting in a high offset frequency noise of about 1250 Hz$^2$/Hz (equivalent short term linewidth of 7.8 kHz and second harmonic linewidth of 31 kHz).
	} 
\end{figure}

\begin{figure}[h]
    \centering
	\includegraphics[width=\linewidth]{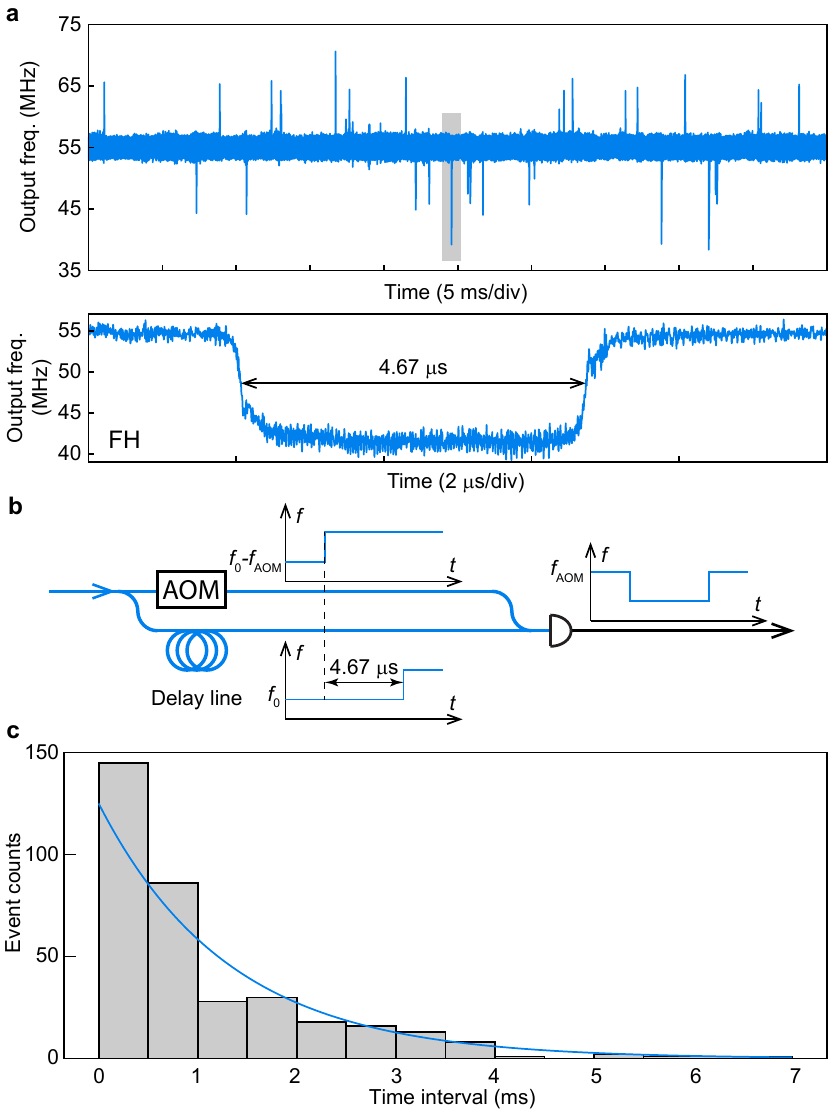}
	\caption{\label{FigS3} Frequency noise measurement in the time domain. {\bf a}, Upper panel: measured output frequency (relative to acousto-optic modulator frequency) of the pump laser in the self-heterodyne setup (see schematic in Fig. 3{\bf a} in main text). Noise spikes appear randomly in the spectrum. Lower panel: zoom-in of the section highlighted in the upper panel showing the shape of a noise spike.  {\bf b}, Simplified schematic of the self-heterodyne setup showing a frequency step event. Upon detection, the time-delayed frequency step forms a square-shaped frequency shift waveform upon. {\bf c}, Measured histogram of the time difference between two frequency step events. A fitted exponential distribution is also plotted.}
\end{figure}

In the Main text, the appearance of increasing low frequency noise in Fig. 3{\bf c} was noted. This noise was observed using several test devices and also using different DFB pump lasers. The magnitude of the noise varied by over an order-of-magnitude. In this section we investigate this noise further by studying frequency noise in the time domain. Specifically, an oscilloscope is used to collect electrical signals from the frequency discriminator outputs (see Fig. 3b lower panel in Main text) instead of directly using a phase noise analyzer. Inspection of the recorded time-domain data provides additional information on frequency discontinuities occurring during the self-injection-locking process as discussed below. For reference, Fig. \ref{FigS2} shows the frequency noise spectra of two SIL systems consisting of different DFB lasers and LN chips. The upper panel data is the same as in the main text, but omits the unfilterd SHG frequency noise spectrum. The high-offset frequency noise levels of the SIL pump (400 Hz$^2$/Hz and 1250 Hz$^2$/Hz) and corresponding short-term linewidth (2.5 kHz and 7.8 kHz) are indicated. The difference in linewidths here is believed to result from differences in LN resonator optical Q factors as well as possible differences in backscattering strength.

The instantaneous frequency of the pump is extracted by applying a fast Fourier transform to the detected photocurrent. An example is plotted in Fig. \ref{FigS3}a. In addition to small frequency fluctuations near the AOM offset frequency of 55 MHz, there are random in time and amplitude noise spikes that appear over the measurement interval. A zoom-in view shows that the spikes have a rectangular shape as plotted in the lower panel in Fig.~\ref{FigS3}{\bf a}. The spikes have a fixed temporal width equal to the delay time of the frequency discriminator (about 4.67 $\mu$s). As the measured frequency is the difference between the laser frequency at a point in time with its later value (delayed by the delay line - see Fig. \ref{FigS3}{\bf b}) each spike can be understood to result from a single frequency step of the laser, where the step amplitude is given by the spike amplitude.

Knowing that each spike has a constant temporal width makes it possible to apply a matched filter to extract the noise spikes from the raw data. Upon extraction from the raw data, the frequency noise for the pump and second harmonic signals are plotted in Fig. \ref{FigS2}a. The filtered data no longer exhibit the rising low-offset-frequency noise, thereby confirming the source of this noise as the observed noise steps. Also notable is that the filtered pump and SHG noise spectra are offset by 6 dB relative to each other, confirming the impact of the SHG process on frequency noise. 

The matched filter approach also enables study of the statistical properties of the noise events. A histogram for the time difference between two frequency steps resembles an exponential distribution (Fig. \ref{FigS3}{\bf c}) and indicates that the underlying process may be Poisson-like, which is expected to produce a $f^{-2}$ dependence in the frequency noise spectrum.  Such a dependence closely matches the measured frequency noise for low offset frequencies in (see dashed line in Fig. \ref{FigS2}a). At this time, the exact cause and dynamics of these steps remains unclear.

Finally, we note that the 6 dB offset in frequency noise between the SIL pump and the SHG signal is also observable in the unfiltered frequency noise spectra provided that the frequency steps do not cause saturation of the homodyne interferometer in the 780 nm band. The second device tested here featured sufficiently lower amplitude steps so that it was possible to make this comparison of the unfiltered data as shown in Fig. \ref{FigS2}b.

\bibstyle{osajnl.bst}
\bibliography{ref_supp.bib}